# Domain wall motion governed by the spin Hall effect

P.P.J. Haazen, E. Murè, J.H. Franken, R. Lavrijsen, H.J.M. Swagten[1], and B. Koopmans

Department of Applied Physics, Center for NanoMaterials and COBRA Research Institute, Eindhoven University of Technology, P.O. Box 513, 5600 MB Eindhoven, The Netherlands

Perpendicularly magnetized materials have attracted tremendous interest due to their high anisotropy, which results in extremely narrow, nano-sized domain walls. As a result, the recently studied current-induced domain wall motion (CIDWM) in these materials promises to enable a novel class of data, memory, and logic devices [1-5]. In this letter, we propose the spin Hall effect as a radically new mechanism for CIDWM. We are able to carefully tune the net spin Hall current in depinning experiments on Pt/Co/Pt nanowires, offering unique control over CIDWM. Furthermore, we determine that the depinning efficiency is intimately related to the internal structure of the domain wall, which we control by small fields along the nanowire. This new manifestation of CIDWM offers a very attractive new degree of freedom for manipulating domain wall motion by charge currents, and sheds light on the existence of contradicting reports on CIDWM in perpendicularly magnetized materials [6-11].

---

[1] h.j.m.swagten@tue.nl

CIDWM is often explained in terms of the adiabatic and nonadiabatic torques [9,10], which both depend on the in-plane spin current that arises from the spin-polarization of the charge current that runs in the ferromagnet. However, in the typical multilayer structures used for domain wall motion in perpendicular materials, a second spin current, generated by the spin Hall effect (SHE) in the adjacent non-magnetic metal layers [14-17], can be injected into the ferromagnet. Materials exhibiting a large SHE are often used for these non-magnetic metal layers, because both the perpendicular anisotropy and the SHE depend strongly on spin-orbit coupling. In such multilayered thin-films, the SHE is particularly efficient in affecting the magnetization because of its large injection interface (the in-plane cross section of the wire). In these structures, spin Hall currents have indeed been shown to change the effective damping [18], induce ferromagnetic resonance [19], inject and detect spin waves [20], and switch the magnetization of in-plane magnetized β-Ta/CoFeB [21] and out-of-plane magnetized Pt/Co/AlO$_x$ [22,23] nanodots. Furthermore, it was suggested that CIDWM in in-plane materials could be influenced [24]. These considerations suggest that the SHE plays an important role in the intensively studied CIDWM in perpendicular materials. In this letter, we explore the potential of CIDWM by the SHE, showing that it in fact constitutes the main contribution to domain wall motion in Pt/Co/Pt.

To study the effect of the spin Hall current on domain wall motion, we have used Pt/Co/Pt structures. Both Pt layers in this stack act as a spin Hall current source, which inject oppositely oriented spins into the ferromagnetic Co layer (see Fig. 1a). Therefore, to inject a net spin current in the Co, the spin Hall currents from the two Pt layers should not cancel fully. This is achieved by choosing unequal Pt layer thicknesses in the range of the spin diffusion length of Pt, since the spin Hall current is dependent on the layer thickness, as was experimentally verified before [19]. Pt/Co/Pt stacks have the further advantage that Rashba effects are negligible. In the closely related Pt/Co/AlO$_x$, in which SHE-induced magnetization reversal has been shown recently [22,23], it was suggested that a nonadiabatic contribution of the Rashba field [25] could be important. In Pt/Co/Pt, there are two approximately equal (Pt/Co) interfaces, resulting in a Rashba field that is negligible (see Supplementary Information for experimental backup). Pt/Co/Pt therefore functions as an excellent model system to unambiguously study the hitherto unexplored role of the SHE in domain wall dynamics.

First, we have verified that the SHE can indeed inject a net spin current, capable of inducing a significant torque on the magnetization, in an asymmetric Pt (4 nm) / Co (0.5 nm) / Pt (2 nm) nanowire. For these unequal Pt layer thicknesses, the net spin Hall current should be approximately 35% of the bulk value [26]. As confirmation, we have performed pure current-induced switching of a uniformly magnetized nanowire, analogous to experiments performed on Pt/Co/AlO$_x$ [22,23]. Current pulses of 5 x 10$^{11}$ A/m$^2$ were injected into a nanowire subjected to an applied field, µ$_0$H$_x$ = 20 mT, parallel to the charge current. Indeed, the current pulses result in magnetization reversal of the nanowire, as is shown in Fig. 1b, where the stable direction of the magnetization is determined by the sign of both the in-plane field and the current, and equal to that observed in Pt/Co/AlO$_x$ [22,23]. This confirms that the thickest Pt layer indeed leads to a larger spin Hall current, resulting in a torque that is not fully compensated by the torque from the spin current from the thinner top layer.

We will now consider the effects of this net spin Hall current on a magnetic domain wall in a nanowire. When injected into a ferromagnetic layer, the spin Hall current gives rise to a torque on the magnetization of the Slonczewski-form [27]. This contribution is added to the LLG equation, which describes the time evolution of the magnetization M:

$$\frac{\partial \vec{M}}{\partial t} = -\gamma \vec{M} \times \vec{H} + \frac{\alpha}{M_s} \vec{M} \times \frac{\partial \vec{M}}{\partial t} - (\vec{u} \cdot \nabla)\vec{M} + \frac{\beta}{M_s} \vec{M} \times (\vec{u} \cdot \nabla)\vec{M} + \frac{\alpha_{SHE}}{M_s} \vec{M} \times (\vec{\sigma} \times \vec{M}), \quad (1)$$

with M$_s$ the saturation magnetization, α the Gilbert damping parameter, γ the gyromagnetic ratio, $\vec{u}$ proportional to the charge current density, and $\alpha_{SHE}$ a parameter determining the strength of the spin Hall effect. The terms on the right hand side denote, in order, precession along an effective field H, the Gilbert damping, the conventional adiabatic and nonadiabatic terms, which we will refer to as gradient torques, and finally the new Slonczewski torque induced by the SHE.

To analyze the effect of the Slonczewksi torque, it is important to consider the internal structure of the domain wall. Since the width of the Pt/Co/Pt nanowires is much larger than the typical domain wall width, the domain walls will be of the Bloch type (see Fig. 1c). For this wall type, the Slonczewksi torque cannot lead to domain wall motion because of symmetry considerations: the 180 degree rotational symmetry around y axis (R$_{2y}$) of this wall type prohibits a well-defined direction of movement, since a hypothetical direction of motion would reverse under this symmetry operation while the system and the resulting Slonczewski torque remain unchanged. However, the Bloch wall can easily be perturbed. In this research, to tune the internal structure of the domain wall, a field in the x direction (along the nanowire) is applied. This applied field changes the domain wall from the initial Bloch type to a Néel type (see Fig. 1c) with the center spin aligned to the field.

When the domain wall is of the Néel type, the spins obtain an x component, which is crucial for the movement of the wall. This dependence on the x component is analogous to the required tilt of the magnetization in the switching that was shown in Figure 1b, where a specific combination of an in-plane field and spin Hall current direction results in a single stable perpendicular magnetization direction. A domain with this magnetization direction is expected to expand under influence of the SHE (Fig. 1d, blue arrows), which we verified by micromagnetic simulations, based on equation (1) (see Supplementary Information). The adiabatic and nonadiabatic gradient terms also give rise to torques on the domain walls, and are expected to push the domain in the electron drift direction, independent on the polarity of the domain [13] (Fig. 1d, yellow arrows). Hence, one domain wall will be driven by a combination of gradient torques and the SHE because they work in parallel for that wall, whereas these two contributions counteract one another in the other domain wall, thereby providing an excellent tool to disentangle these contributions to the CIDWM.

This scheme to uniquely identify the role of the SHE is now applied to a Pt (4nm) / Co (0.5 nm) / Pt (2 nm) nanowire, in which a well-defined region with reduced anisotropy is engineered using $Ga^+$ ion irradiation [28]. In this magnetically softer region, a domain can be stabilized, as can be seen in Fig. 2a (left pane). Since the energy of a domain wall scales with the root of the anisotropy, the domain walls stay pinned at the anisotropy steps. When the perpendicular $H_z$ field is increased, a critical field $H_{depin}$ will depin the domain walls over the energy barriers, after which they propagate towards the ends of the nanowire.

We will now concentrate on the dependence of $H_{depin}$ on an in-plane current for both domain walls where we set the Néel structure by applying an in-plane field $\mu_0 H_x$ = -15 mT. As can be seen in Fig. 2b, already at reasonably low current density J (i.e., ~ $10^{10}$ A/m$^2$), $H_{depin}$ can be significantly altered, and an almost linear dependence on the current is measured. However, the observed symmetry is radically different from that expected of the conventional gradient torques. When the center domain is magnetized in the upward direction and a negative $H_x$ is applied (Fig. 2b), a positive current results in a lower $H_{depin}$ for both domain walls, and is therefore assisting the depinning of both domain walls, even though their depinning directions are opposite. When the polarity of the domain wall is reversed (Fig. 2c), the required $|H_{depin}|$ increases with positive current, and the current now opposes the DW depinning, again for both domain walls. When the in-plane field is reversed, the slopes of $H_{depin}$ versus J change sign, as can be seen in Fig. 2d and Fig. 2e. Such behavior cannot be explained in the conventional paradigm of gradient torques, which predicts an opposite sign of the slopes for the two domain walls, and none of the observed sign changes. Moreover, we find no systematic difference in the slopes for the two domain walls, indicating that the conventional gradient torques are negligible. Instead, we have demonstrated a new mechanism for domain wall motion, governed by the spin Hall effect. Note that the observed dependence of the depinning field at zero current on $H_x$ and the perpendicular magnetization direction, visible as the offsets in Fig. 2 b-e, is caused by the local energy landscape of the depinning center, thereby forming a separate effect that is not relevant for the SHE behavior of $dH_{depin}/dJ$ (see Supplementary Information).

To further proof that the spin Hall effect is of dominant importance for the domain wall depinning, we will now discuss the role of the stack composition. The subtractive nature of the two competing spin currents from the Pt layers predicts that engineering of the strength and sign of CIDWM by tuning the Pt thicknesses is possible. Therefore, we have repeated these measurements on Pt (x nm) /Co (0.5 nm)/Pt (y nm) nanowires with (x,y)=(4,2);(3,3);(2,4). Indeed, the sign of the depinning efficiency, $\epsilon = \mu_0 \times dH_{depin}/dJ$, clearly reverses between the (4,2) and (2,4) stacks, as can be seen in Figure 3. Furthermore, for the (3,3) stack, the two spin currents cancel, resulting in zero net spin current and no systematic influence on the domain wall depinning.

The functional dependence of $\epsilon$ on $H_x$ also shows an interesting behavior. After a steep increase, at $\mu_0 H_x$ > 15 mT, $\epsilon$ levels off for both the (4,2) and (2,4) stacks. Apart from predicting the observed linear dependence of $H_{depin}$ on current, micromagnetic simulations using only the spin Hall induced Slonczewski torque also reproduce this saturation, without using any free parameters (see Supplementary materials), as can be seen in Fig. 3 (dashed lines). They reveal that the internal structure of the domain wall is indeed of crucial importance for $\epsilon$. At $H_x=0$, the domain wall is of the Bloch type, and the depinning efficiency is zero. When an $H_x$ is applied, the internal angle of the domain wall starts to align with this field, and $\epsilon$ increases. At $\mu_0|H_x| \cong 15$ mT, the domain wall is fully aligned with $H_x$ (i.e. in a full Néel configuration, see Supplementary Information), and $\epsilon$ saturates. These results show that it is possible to tune the efficiency and the direction of the CIDWM by controlling the magnitude of the net spin Hall current and the internal domain wall structure.

The findings presented in this letter have important implications for the research field, where the spread in sign and magnitude in reported values of $\epsilon$ is an urgent issue [6-11]. Because of the abundant use of materials with high spin orbit coupling in perpendicularly magnetized domain wall conduits, it is very likely that the SHE also plays a role in other CIDWM experiments. Even without the use of an in-plane field, deviations from a pure Bloch structure can be induced by other factors, such as field misalignments or contributions from the adiabatic and nonadiabatic torques, or from the shape anisotropy in narrow wires. Hence, it is likely that in previous research the SHE has influenced the CIDWM, and these contributions could have been erroneously ascribed to the nonadiabatic torque. We therefore belief that the SHE plays a decisive role in explaining, at least partially, the existence of contradicting reports on CIDWM in perpendicular materials.

Finally, for domain-wall based applications, the demonstration of the SHE driven domain wall motion offers a completely new degree of freedom for controlling domain wall motion by a charge current. We have shown that when the domain wall structure is controlled, reliable SHE driven domain wall motion can be achieved. In narrow wires (<~ 60 nm [29]), the shape anisotropy favors domain walls of the Néel type, which would allow for spin Hall induced domain wall motion without the need for applied in-plane fields in these wires, where the initial configuration of the domain wall can be tuned to set the direction of motion. This favorable scaling behavior makes the SHE driven domain wall motion especially promising, since it opens up possibilities for efficient and dense data storage devices.

## Methods

The dimensions of the Pt/Co/Pt nanostrips are 1.5 µm x 20 µm x 6.5 nm. Pt and Co were deposited on thermally oxidized $SiO_x$ substrates by DC magnetron sputtering in a system with a base pressure of ~3x10$^{-8}$ mbar. From these thin films, nanostrips were fabricated using e-beam lithography and lift-off. The electrodes were made of 35 nm thick Pt and were also deposited by sputter deposition. The out-of-plane component of the magnetization ($M_z$) of the nanostrips was measured by polar Kerr microscopy. The external magnetic field was applied in 3 orthogonal directions. The $H_x$ field was applied before nucleation, and kept constant until completion of the measurement routine. No contributions from Joule heating, which would have resulted in a deviation from the linear behavior of $H_{depin}$ versus J, can be observed in the depinning experiments, because low current densities (< 2.5 x 10$^{10}$ A/m$^2$) were used.

## Acknowledgements

The work is part of the research programme of the Foundation for Fundamental Research on Matter (FOM), which is part of the Netherlands Organisation for Scientific Research (NWO). E.M. acknowledges support from the Swiss National Science Foundation (SNSF), Grant No. PBELP2-130894.

## Author contributions

P.P.J.H., E.M., J.H.F. and R.L. designed the experiment. P.P.J.H. carried out the experiment and the simulations and prepared the manuscript with help from E.M. and J.H.F.; H.J.M.S. and B.K. supervised the study. All authors discussed the results and commented on the manuscript.

**Bibliography**

[1] Hayashi, M., Thomas, L., Moriya, R., Rettner, C. & Parkin, S. S. P. Current-controlled magnetic domain-wall nanowire shift register. *Science* **320,** 209-211 (2008).

[2] S. S. P. Parkin, U.S. Patent 6834005 (2004).

[3] Honjo, H. et al. Domain-wall-motion cell with perpendicular anisotropy wire and in-plane magnetic tunneling junctions. *J. Appl. Phys.* **111,** 07C903 (2012).

[4] Allwood, D. A. et al. Magnetic domain wall logic. *Science* **309,** 1688–1692 . (2005).

[5] Zhang Y., Zhao, W. S., Ravelosona, D., Klein, J.-O., Kim, J. V. & Chappert, C. Perpendicular-magnetic-anisotropy CoFeB racetrack memory. *J. Appl. Phys.* **111,** 093925 (2012).

[6] Miron, I. M. et al. Domain Wall Spin Torquemeter. *Phys. Rev. Lett.* **102,** 1-4. (2009).

[7] Moore, T. A. et al. High domain wall velocities induced by current in ultrathin Pt/Co/AlOx wires with perpendicular magnetic anisotropy. *Appl. Phys. Lett.* **93,** 262504 (2008).

[8] Miron, I. M. et al. Fast current-induced domain-wall motion controlled by the Rashba effect. *Nature Materials* **10,** 419-423 (2011).

[9] Kim, K.-J. et al. Electric Control of Multiple Domain Walls in Pt/Co/Pt Nanotracks with Perpendicular Magnetic Anisotropy *Appl. Phys. Express* **3,** 083001 (2010).

[10] Heinen, J. et al. Current-induced domain wall motion in Co/Pt nanowires: Separating spin torque and Oersted-field effects. *Appl. Phys. Lett.* **96,** 202510 (2010).

[11] Lavrijsen, R. et al. Asymmetric Pt/Co/Pt-stack induced sign-control of current-induced magnetic domain-wall creep. *Appl. Phys. Lett.* **100,** 262408 (2012).

[12] Zhang, S. & Li, Z. Roles of Nonequilibrium Conduction Electrons on the Magnetization Dynamics of Ferromagnets. *Phys. Rev. Lett.* **93,** 1-4 (2004).

[13] Thiaville, A, Nakatani, Y., Miltat, J. & Suzuki, Y. Micromagnetic understanding of current-driven domain wall motion in patterned nanowires. *Europhys. Lett.* **69,** 990-996. (2005).

[14] Dyakonov, M. I. & Perel, V. I. Possibility of orienting electron spins with current. *JETP Lett.* **13,** 467-469 (1971).

[15] Dyakonov, M. I. & Perel, V. I. Current-induced spin orientation of electrons in semiconductors. *Phys. Lett. A* **35,** 459-460 (1971).

[16] Hirsch, J. E. Spin Hall effect. *Phys. Rev. Lett.* **83,** 1834-1837 (1999).

[17] Kato, Y. K., Myers, R. C., Gossard, a C. & Awschalom, D. D. Observation of the spin Hall effect in semiconductors. *Science* **306,** 1910-1913 (2004).

[18] Demidov, V. E., Urazhdin, S., Edwards, E. R. J. & Demokritov, S. O. Wide-range control of ferromagnetic resonance by spin Hall effect. *Appl. Phys. Lett.* **99,** 172501 (2011).

[19] Liu, L., Moriyama, T., Ralph, D. C. & Buhrman, R. A. Spin-Torque Ferromagnetic Resonance Induced by the Spin Hall Effect. *Phys. Rev. Lett.* **106,** 036601 (2011).

[20] Kajiwara, Y. et al. Transmission of electrical signals by spin-wave interconversion in a magnetic insulator. *Nature* **464,** 262-266. (2010).

[21] Liu, L. et al. Spin-Torque Switching with the Giant Spin Hall Effect of Tantalum. *Science* **336,** 555-558 (2012).

[22] Miron, I. M. et al. Perpendicular switching of a single ferromagnetic layer induced by in-plane current injection. *Nature* **476,** 189-193 (2011).

[23] Liu, L., Lee, O. J., Gudmundsen, T.J., Ralph, D. C., & Burhman, R. A. Current-Induced Switching of Perpendicularly Magnetized Magnetic Layers Using Spin Torque from the Spin Hall Effect. *Phys. Rev. Lett* **109**, 096602 (2012)

[24] Seo, S.-M., Kim, K.-W., Ryu, J., Lee, H.-W. & Lee, K.-J. Current-induced motion of a transverse magnetic domain wall in the presence of spin Hall effect. *Appl. Phys. Lett.* **101**, 022405 (2012)

[25] Wang, X. & Manchon, A. Diffusive Spin Dynamics in Ferromagnetic Thin Films with a Rashba Interaction. *Phys. Rev. Lett.* **108,** 1-5 (2012).

[26] Liu, L., Burhman, R. A. & Ralph, D. C. Review and Analysis of Measurements of the Spin Hall Effect in Platinum. Preprint at <http://arxiv.org/abs/1111.3702v3> (2012).

[27] Slonczewski, J. Current-driven excitation of magnetic multilayers. *J. Magn. Magn. Mater.* **159,** L1-L7 (1996).

[28] Franken, J. H. et al. Precise control of domain wall injection and pinning using helium and gallium focused ion beams. *J. Appl. Phys.* **109,** 07D504 (2011).

[29] Koyama, T. et al. Observation of the intrinsic pinning of a magnetic domain wall in a ferromagnetic nanowire. *Nature materials* **10,** 194-197 (2011).
10

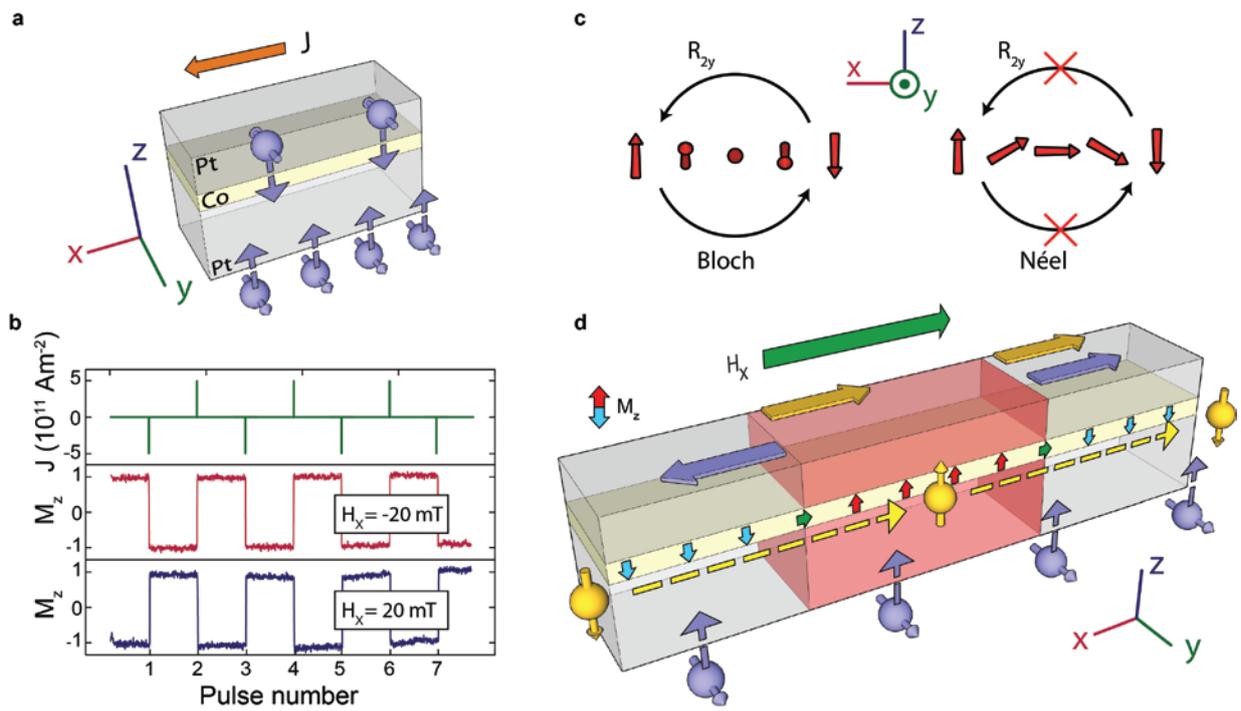

**Figure 1 | Magnetization dynamics induced by the Spin Hall effect. a,** A vertical spin current is generated in both Pt layers as a consequence of the charge current density J via the SHE and injected in the Co. The thickest Pt layer induces a higher spin current, leading to a nonzero net injected spin current. **b**, Perpendicular switching of a uniformly magnetized Pt (2 nm) / Co (0.5 nm) / Pt (4 nm) nanowire confirms the torque from the spin Hall current, where the combination of the charge current direction and the in-plane field $H_x$ set the stable perpendicular magnetization direction. **c**, Bloch and Néel domain wall types. The Bloch wall is symmetric under a 180 degree rotation along the y axis ($R_{2y}$), which prohibits motion of the domain wall when subjected to a Slonczewski torque. **d**, Contributions to the CIDWM (arrows on top of structure) from conventional gradient torques (yellow) and the spin Hall effect (violet). For simplicity, only the dominating spin current from the bottom Pt layer is visualized. Under the application of an applied magnetic field in the x direction ($H_x$), the Néel wall can be stabilized, with its center spin pointing along the field.

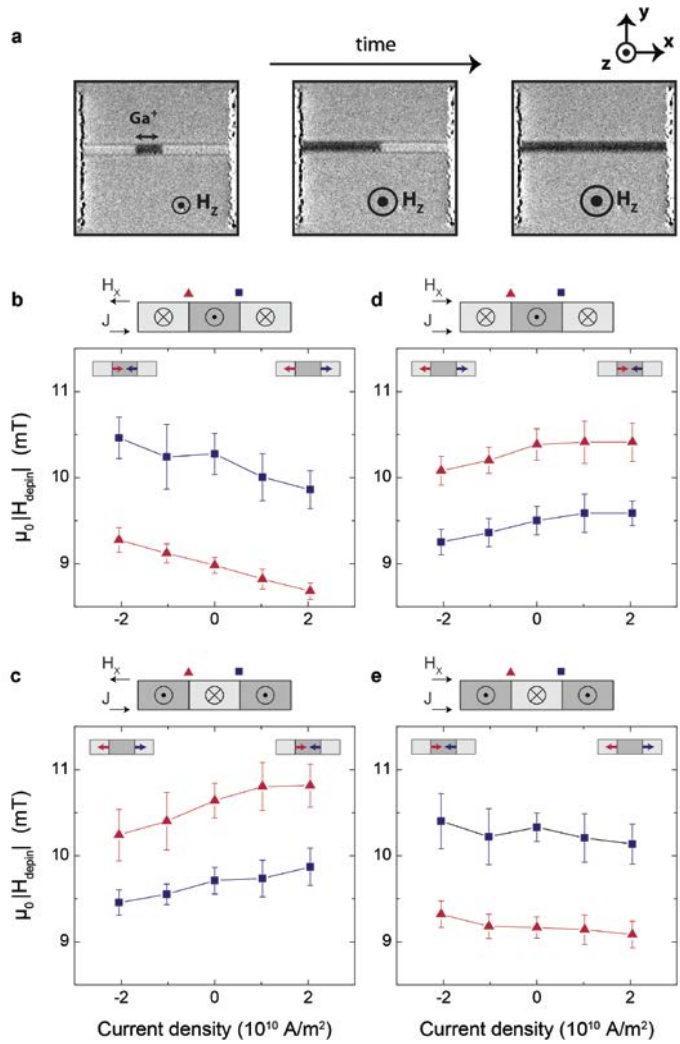

**Figure 2 | Domain wall depinning experiment. a** Polar Kerr images of the subsequent nucleation of a domain in the low-anisotropy region and the depinning events of the two walls of the domain. **b-e** Depinning fields for the right (blue squares) and left (red triangles) domain walls versus in-plane current in Pt (4 nm) / Co (0.5 nm) / Pt (2 nm). Data points are averaged values of 20 depinning events, with the standard deviation given by the error bars. Measurements were performed with in-plane fields of $\mu_0 H_x = -15$ mT (**b,c**) and $\mu_0 H_x = 15$ mT (**d,e**). The sign of the contribution of the current to the domain wall depinning changes under reversal of domain wall polarity (see **b** vs **c** and **d** vs **e**) and in-plane field (**b** vs **d** and **c** vs **e**). The cartoons on top of the graphs shows the labels of the two domain walls as a legend, and the cartoons in the graphs show the direction of the current induced contribution to the depinning process for negative (left side) and positive (right side) currents.

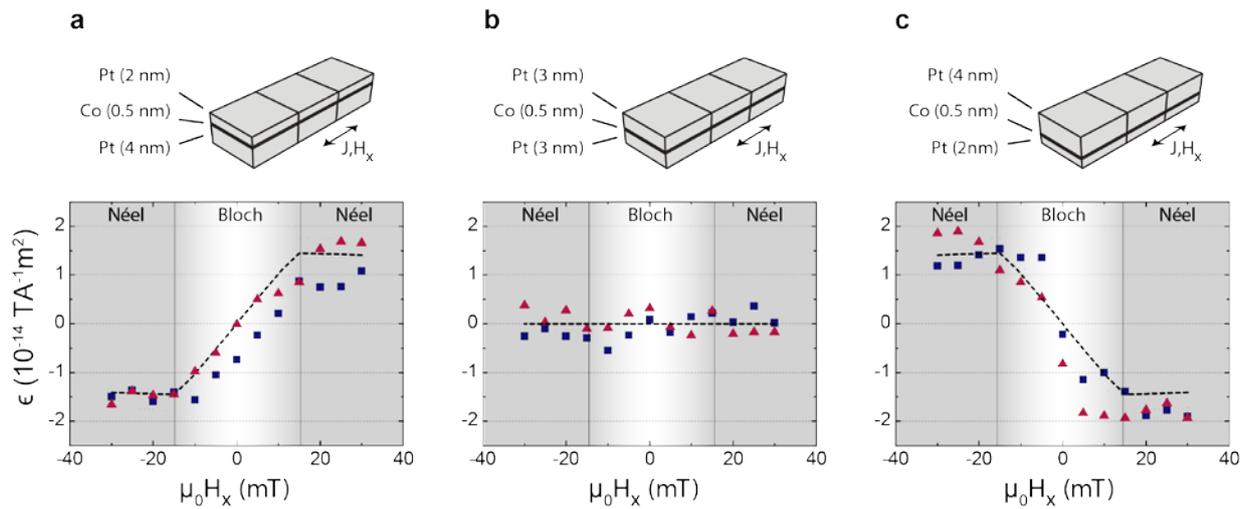

**Figure 3 | Depinning efficiency as a function of Hx field, for Pt (x nm)/Co (0.5 nm)/Pt (y nm).** **a** (x,y) = (4,2); **b** (x,y) = (3,3); **c** (x,y) = (2,4). The magnetization of the expanding domain is parallel to +z. The cartoons on top of the graphs indicate the stack sequence. The dashed black lines are the results of the micromagnetic simulations, performed without adjustable parameters (see Supplementary Information), and the symbols indicate $\epsilon$ for the two domain walls, with the color coding analogous to Fig. 2. The internal structure of the domain wall, as determined by micromagnetic simulations, is indicated by the background color of the graph. At fields higher than $\mu_0 H_x = 15$ mT, the structure is of the Néel type, and at $\mu_0 |H_x| < 15$ mT, the domain wall structure changes from the Néel type to the Bloch type (at $\mu_0 H_x = 0$ mT) to the opposite Néel type.

**SUPPLEMENTARY INFORMATION**

**S1. Predicted ϵ from the nonadiabatic torque**

In the main text, we have shown that the conventional gradient torques are of negligible influence on the depinning. However, the role of both the adiabatic and non-adiabatic spin transfer torques has been well-established, especially in in-plane materials, and the present research does not exclude the existence of these torques. Here, we will briefly discuss the expected influence of the nonadiabatic torque, since high domain wall velocities are usually ascribed to this torque.

ϵ can be related to β via the 1D framework developed by Thiaville *et al.* [S1]:

$$\epsilon = \frac{\beta P \hbar}{2 e M_S \Delta}$$

Which yields ϵ ≈ 0.75β, using P=0.4, $M_s$=1.4e6 MAm$^{-1}$, Δ=1.5e-8 m, A=6.5 x 10$^{-9}$ * 1.5 x 10$^{-6}$ m$^2$. Here, we have overestimated the expected value of β, since a constant current density is assumed (~8% through Co), whereas the current density in Co is expected to be lower (~3%) [S2]. Since the dynamics of both domain walls differ only by the sign of their magnetization gradient, the contribution of β to $dH_{depin}/dI$ can be estimated by the difference in responses ($\epsilon_{DW1}$ and $\epsilon_{DW1}$) of the 2 domain walls:

$$\epsilon_\beta = \frac{\epsilon_{DW1} - \epsilon_{DW2}}{2}$$

No consistent $\epsilon_\beta$ was found in our data. However, $\epsilon_\beta$ of the order 0.1 (β≈0.13) would fall within the uncertainty of our experiment.

**S2. Effects of the local crystalline structure**

As was shown in the main text (Fig. 2), at zero current, the depinning field is dependent on both $H_x$ and the perpendicular magnetization direction. This behavior is caused by the local energy landscape of the depinning center, and differs per pinning site and nanowire in a nonsystematic way. Fig. S1 shows the observed values of $H_{depin}$ as a function of $H_x$, without applied current, measured on two nominally equal

Pt (2 nm) / Co (0.5 nm) / Pt (4 nm) nanowires that were grown in one run on the same sample. As can be seen, the observed behavior of $H_{depin}$ as a function of $H_x$ does not reproduce, and no correlation can be found between depinning sites.

Note that, in a sample without disorder or defects, $|H_{depin}|$ at zero current should have even more symmetry than is observed in Fig. S1: It should be an even function of $H_x$, and independent of the perpendicular magnetization direction of the domain. In the experimental data, there is an additional symmetry breaking factor that breaks the symmetry under 180° rotation along the z axis and 180° rotation along the x axis. This symmetry breaking can be explained by considering that the ferromagnetic layers, although macroscopically symmetric under these rotations, have an internal defect structure, which influences the magnetization, and can therefore explain the observed symmetry in our results. On the other hand, the behavior of ϵ is highly reproducible, as one can readily see in the similar behavior of the two domain walls in Fig. 3 of the main text.

## S3. Micromagnetic simulations

To study the depinning behavior of a domain wall subjected to a Slonczewski torque, we have performed 1D micromagnetic simulations, based on equation (1). In this section, we will present more details on these simulations.

The extended LLG equation (equation (1)) has been solved on a 1D grid of 100 sites with 4 nm spacing. The exchange interactions and perpendicular anisotropy are included as effective fields. Since the gradient torques do not give a significant contribution to ϵ in our experiments, they are not included in the simulations. The demagnetization fields in the x direction are determined at each timestep by calculating the field of the magnetic surface charges between the sites. The demagnetization fields in the y direction were neglected, since a very wide nanowire was assumed. The demagnetization fields in the z direction are included as a correction to the perpendicular anisotropy, which holds in the limit of an infinitely thin film. To check the validity of our simulations, we have verified that we can reproduce the standard domain wall profiles, as well as the conventional current-driven behavior by gradient torques, including the Walker breakdown transition.

The strength of the effective perpendicular anisotropy, which is the perpendicular anisotropy (uniaxial anisotropy with $K_u$ = 1.5 x $10^6$ $Jm^{-3}$) compensated for the perpendicular demagnetization fields, was lowered on one side of the grid to simulate the $Ga^+$ irradiated region, as can be seen in Figure S2b. On the right side, the effective anisotropy is unchanged, with $K_{eff} \cong$ 0.268 $MJm^{-3}$. On the left side, the anisotropy is decreased by a factor $\Delta K$, so that the effective anisotropy on the left side equals $K_{eff,Ga+}$ = (1- $\Delta K$) $K_{eff}$.

To calculate the depinning field for a fixed $H_x$ and Slonczewski torque, the domain wall is first allowed to relax without a perpendicular $H_Z$ field for 30 ns. Then, $H_Z$ is incremented in steps of $10^{-5}$ T, and the magnetization is again allowed to relax for 15 ns in between these increments. After each increment of $H_Z$, the state of the spins in the right region is inspected. If the perpendicular component of the magnetization direction is reversed in this region, the current $H_z$ field is saved as the depinning field $H_{depin}$. This procedure is repeated for different values of $H_x$ and Slonczewski torque magnitude.

To calculate the net spin Hall current, we have used the spin Hall angle of Pt, $\theta_{SH}$ = 0.068 from ref [S3]. In the same article, the thickness dependence of the spin Hall current on the Pt layer thickness is experimentally studied in the range 2-10 nm, from which one can deduce that the net spin Hall current is only 35% of the bulk value for Pt/Co/Pt stacks with 2 nm and 4 nm thick Pt layers. The resulting torque is added to the LLG equation as described by equation (1) (see main text), with the prefactor:

$$\alpha_{SHE} = 0.35 \frac{\gamma \hbar \theta_{SH}}{2eM_S t \mu_0} J,$$

with e the electron charge, t the Co layer thickness, $\mu_0$ the permeability of free space and J the current density and $\hbar$ the reduced Planck's constant. The other simulation parameters that were used are the exchange stiffness constant A = 1.6 x $10^{-11}$ $Jm^{-1}$, the saturation magnetization $M_s$ = 1.4 x $10^6$ $Am^{-1}$, and a Gilbert damping constant of $\alpha$=0.1.

Analogous to the experimental results, the calculated $H_{depin}$ has a linear dependence on the current density, as can be seen in Fig. S2c, where $H_{depin}$ is plotted as a function of the current density at an in-plane field of $\mu_0 H_x = 20$ mT. Finally, in Fig. S2d, the dependence of $\epsilon$ versus the in-plane field is depicted. A gradual change in the internal angle of the domain wall was visible in the simulations. At $\mu_0 H_x = 0$ mT, the domain wall is of the Bloch type, and there is no influence of the current on $H_{depin}$. At higher $H_x$, the internal angle of the domain wall aligns with the $H_x$, and at approximately $\mu_0|H_x| = 15$ mT, the Néel wall is stabilized and $\epsilon$ levels off. To see the correspondence between $\epsilon$ and the domain wall structure, in the bottom pane of Fig. S2d, the internal angle of the domain wall, without a spin Hall current or anisotropy step, is plotted as a function of $H_x$. Furthermore, the calculations have been repeated without the in-plane demagnetization fields, which leads to a step in $\epsilon$ as a function of $H_x$, because the domain wall is now free to align with the applied $H_x$ field.

To study the influence of the anisotropy step height $\Delta K$, the calculations have been repeated for multiple step heights (open diamonds and solid squares in Fig. S2d). As can be seen in Fig. S2d, the behavior of $\epsilon$ is almost not influenced by the step height.

As can be seen in Fig. 3 of the main text, these simulations match quantitatively with the experimentally observed values for $\epsilon$, although no thermal and structural disorder is included and the step and the anisotropy gradient is approximated by a step function. Although these simplifications prohibit us from accurately approximating $H_{depin}$, the current-field equivalency $\epsilon$ does only take into account the magnitude of the contributions of $H_z$ and the spin Hall current and can therefore be approximated with this relatively simple model.

**S4. Magnitude of Rashba contributions**

In Pt/Co/AlO$_x$, a strong Rashba field in the *y* direction ($H_{Rashba}$) can be induced by a current running in the *x* direction, due to the lack of structural inversion symmetry. This field scales linearly with the current, with a prefactor $H_{Rashba}/J$ of $1\pm0.1 \times 10^{-12}$ Tm$^2$A$^{-1}$ [S4]. Nonadiabatic contributions from $H_{Rashba}$ have been predicted to have the same symmetry as the Slonczewski torque from the SHE, although such effects have never been measured in a separate experiment before. In Pt/Co/AlO$_x$, there has been a discussion on whether these nonadiabatic contributions of H$_{Rashba}$ could contribute to the switching that was observed with an H$_x$ field and an in-plane current [S5]. Here, we will show experimentally that these contributions are very small and that Rashba effects are negligible in our structures.

The existence of a net H$_{Rashba}$ can only arise in structurally asymmetric layers. In Pt/Co/Pt structures, the Co is enclosed between two similar interfaces, and no significant influences from a net Rashba field are expected. Indeed, the existence of a net Rashba field in Pt (3nm) / Co (0.6 nm) / Pt (3nm) nanowires was already refuted by Miron et al. [S4]. Small Rashba effects could in principle arise from second-order contributions, such as growth-related differences in the interface or unequal current densities at the interfaces [S6].

To study the magnitude of H$_{Rashba}$ due to these effects in Pt/Co/Pt structures with unequal Pt thickness, we have studied the dependence of the switching field H$_{switch}$ on the in-plane field H$_y$ for different current directions. Generally, the switching field of perpendicularly magnetized materials is lowered when in-plane fields are applied, an effect which was also exploited by Miron *et al.* When an H$_{Rashba}$ is induced, it will enhance or oppose the applied H$_y$ field, resulting in a shift in the dependence of H$_{switch}$ on H$_y$.

Figure S3 shows the dependence of H$_{switch}$ as a function of H$_y$ for a Pt (2nm) / Co (0.5 nm) / Pt (4nm) nanowire. Note that the current densities here are ~20 x higher than used to study the domain wall depinning. Still, the observed shift between the positive and negative current direction is very small, 5 ± 3 mT, corresponding to a prefactor H$_{Rashba}$/J ≈ 1.2 x 10$^{-14}$ T/m$^2$, almost two orders of magnitude lower than that observed in Pt/Co/AlO$_x$ [S4]. Furthermore, these current-induced fields in the *y* direction can also be explained by considering the Oersted fields in our structure. Since the magnetization reversal (Fig. 1 of main text) in our Pt/Co/Pt structures can be achieved with current densities of the same order of magnitude as in Pt/Co/AlO$_x$, the contribution from (nonadiabatic) Rashba effects is negligible.

**S5. Magnetization reversal of (4,2) and (2,4) stacks**

The magnetization reversal experiments on Pt (4 nm) / Co (0.5 nm) / Pt (2 nm) nanowires have been repeated on Pt (2 nm) / Co (0.5 nm) / Pt (4 nm) and Pt (3 nm) / Co (0.5 nm) / Pt (3 nm). For the Pt (3 nm) / Co (0.5 nm) / Pt (3 nm), no switching was observed, which we ascribe to an almost perfect cancellation of the two spin currents. The Pt (2 nm) / Co (0.5 nm) / Pt (4 nm) does switch under the application of current pulses, with a sign of the stable magnetization direction opposite tot that observed for the Pt (4 nm) / Co (0.5 nm) / Pt (2 nm) stack, as can be seen Fig. S4.

**Bibliography Supplementary Materials**

[S1] A. Thiaville, Y. Nakatani, J. Miltat, Y. Suzuki, *Europhys. Lett.* **69,** 990 (2005)

[S2] Cormier, M., Mougin, A., Ferré, J., Thiaville, a., Charpentier, N., Piéchon, F., Weil, R., et al. Effect of electrical current pulses on domain walls in Pt/Co/Pt nanotracks with out-of-plane anisotropy: Spin transfer torque versus Joule heating. *Physical Review B* **81**, 1-9 (2010)

[S3] Liu, L., Buhrman, R.A., and Ralph, D.C. Review and Analysis of Measurements of the Spin Hall Effect in Platinum, *ArXiv* 1111.3702v3.

[S4] Miron, I. M., Gaudin, G., Auffret, S., Rodmacq, B., Schuhl, A., Pizzini, S., Vogel, J., et al. Current-driven spin torque induced by the Rashba effect in a ferromagnetic metal layer. *Nature materials* **9**, 230-234 (2010)

[S5] Miron, I. M., Garello, K., Gaudin, G., Zermatten, P.-J., Costache, M. V., Auffret, S., Bandiera, S., et al. Perpendicular switching of a single ferromagnetic layer induced by in-plane current injection. *Nature* **476**, 189-193 (2011)

[S6] Lavrijsen, R., Haazen, P. P. J., Murè, E., Franken, J. H., Kohlhepp, J. T., Swagten, H. J. M., Koopmans, B. Asymmetric Pt/Co/Pt-stack induced sign-control of current-induced magnetic domain-wall creep. *Applied Physics Letters*, 100(26), 262408. (2012)

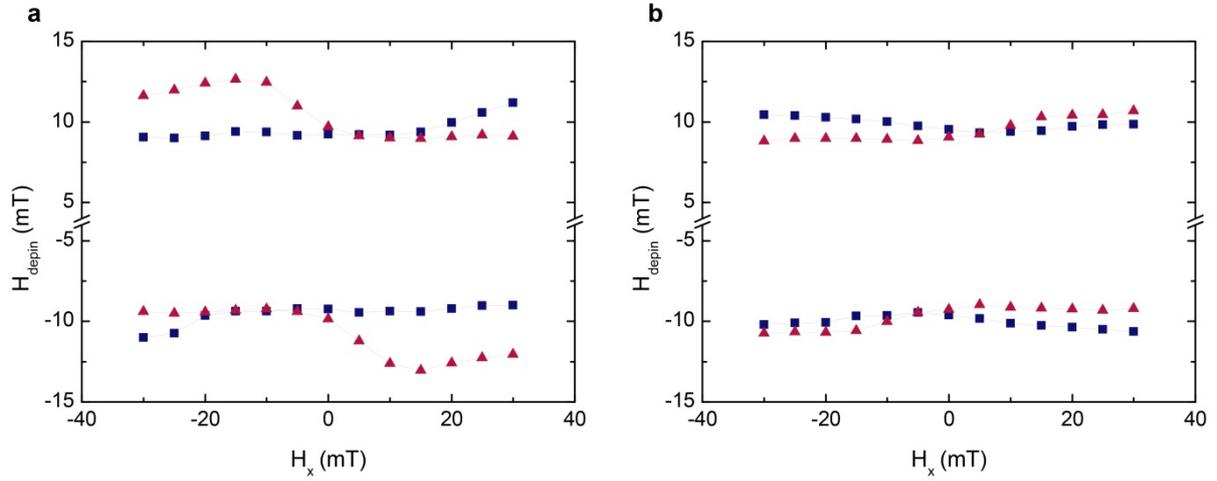

**Figure S1 | Typical measurements of depinning field without current as a function of $H_x$. a** and **b** are two nanowires on the same Pt (2 nm) / Co (0.5 nm) / Pt (4 nm) sample. The red triangles and the blue squares denote the two walls of a single domain, analogous to Fig. 2 of the main text. No correlation between nanowires or depinning sites in the behavior of $H_{depin}$ without current as a function of $H_x$ could be found; the dependence is governed by the local defect structure near the depinning centers.

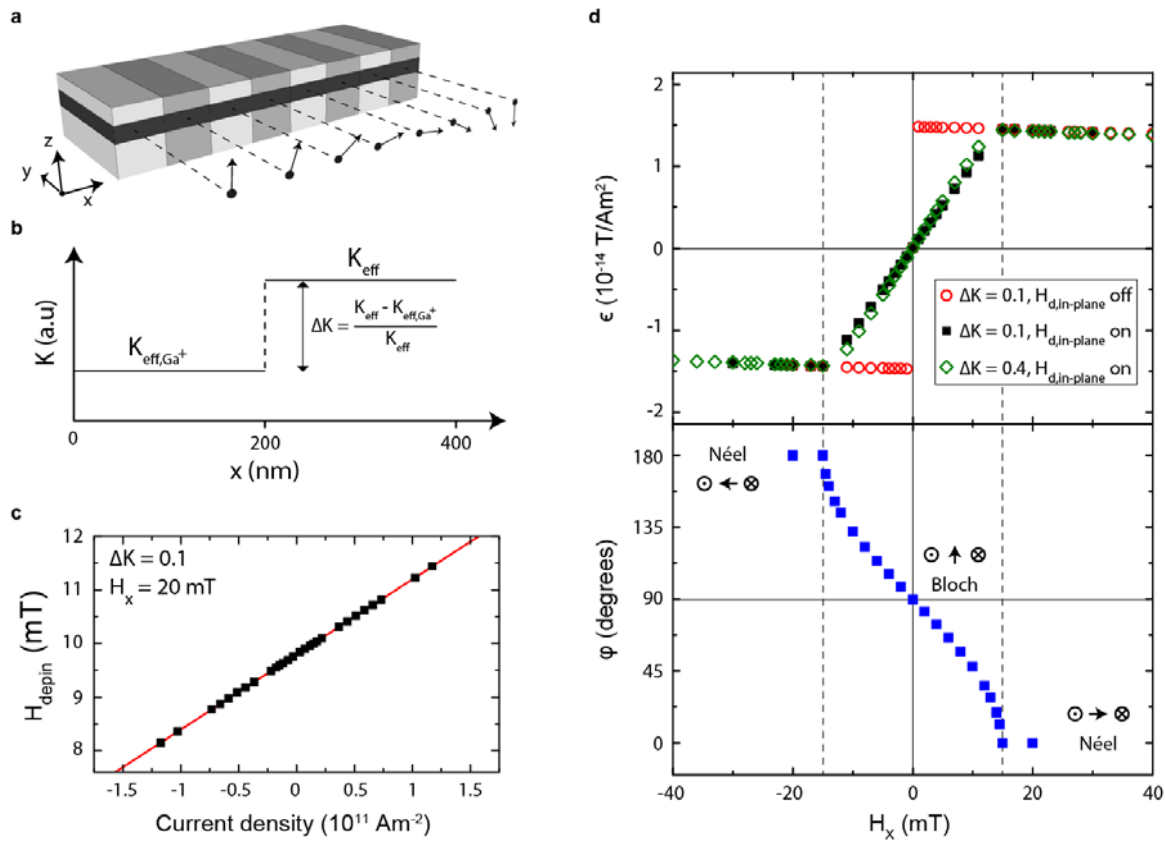

**Figure S2 | ϵ as a function of $H_x$ as determined by micromagnetic simulations. a** Cartoon of the discretization of the magnetization, used for the micromagnetic simulations of the depinning behavior; **b** The change in anisotropy is modeled as a step in the effective anisotropy $K_{eff}$ at the center of the simulation region. **c** Simulation results of the dependence of $H_{depin}$ on the current density in a Pt (4 nm)/Co (0.5 nm)/Pt (2 nm) film. The red line is a linear fit of the data points. **d** Results of micromagnetic simulations: in the top graph, ϵ is plotted as a function of $H_x$, for two step heights and once without the in-plane demagnetization fields. The bottom graph shows the in-plane angle of a domain wall in equilibrium without SHE or anisotropy steps.

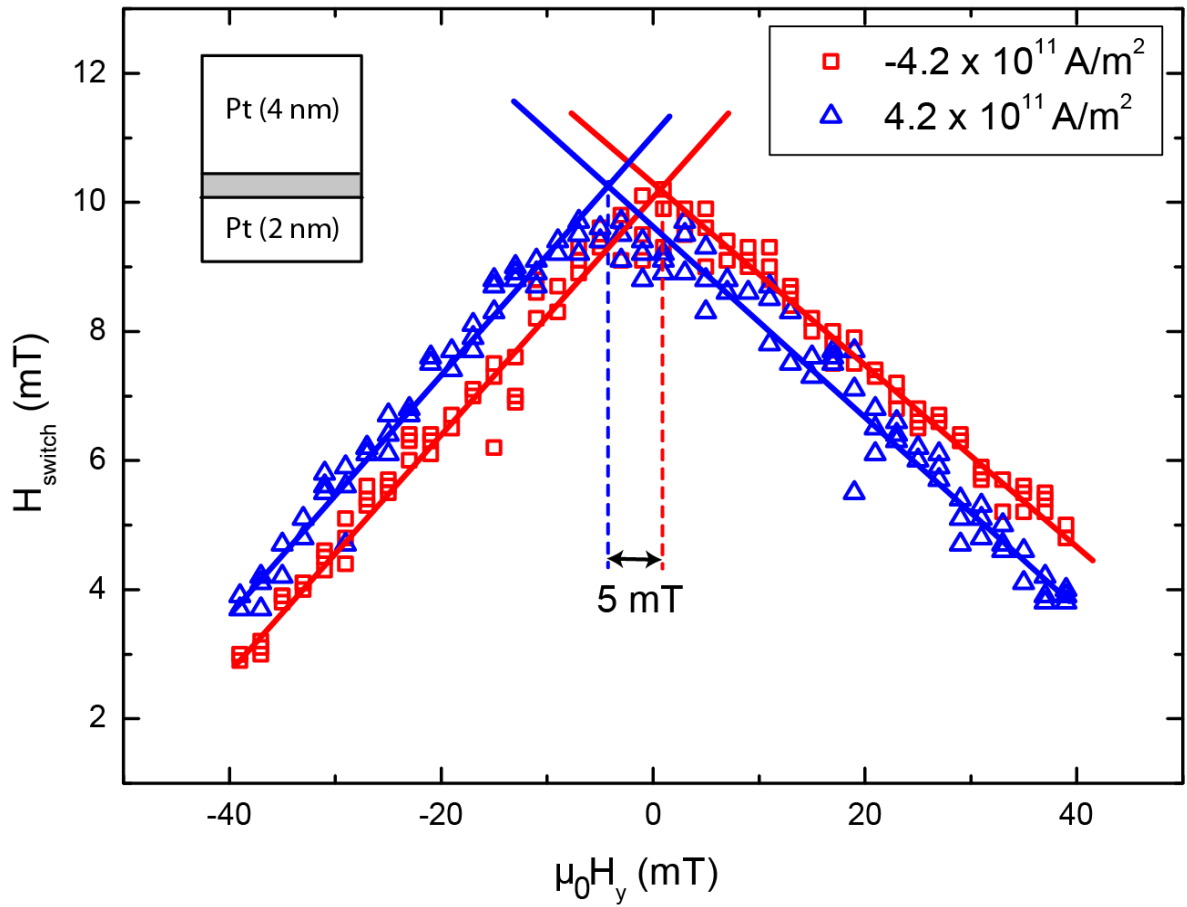

**Figure S3 | Switching field for Pt (2nm) / Co (0.5 nm) / Pt (4 nm) wire as a function of in-plane current.** Inverting the sign of the current yields only a 5±3 mT shift in the graphs, indicating that the $H_y$ field generated by the current is much lower than that for typical Rasbha fields observed in Pt/Co/AlO$_x$. The lines are guides to the eye.

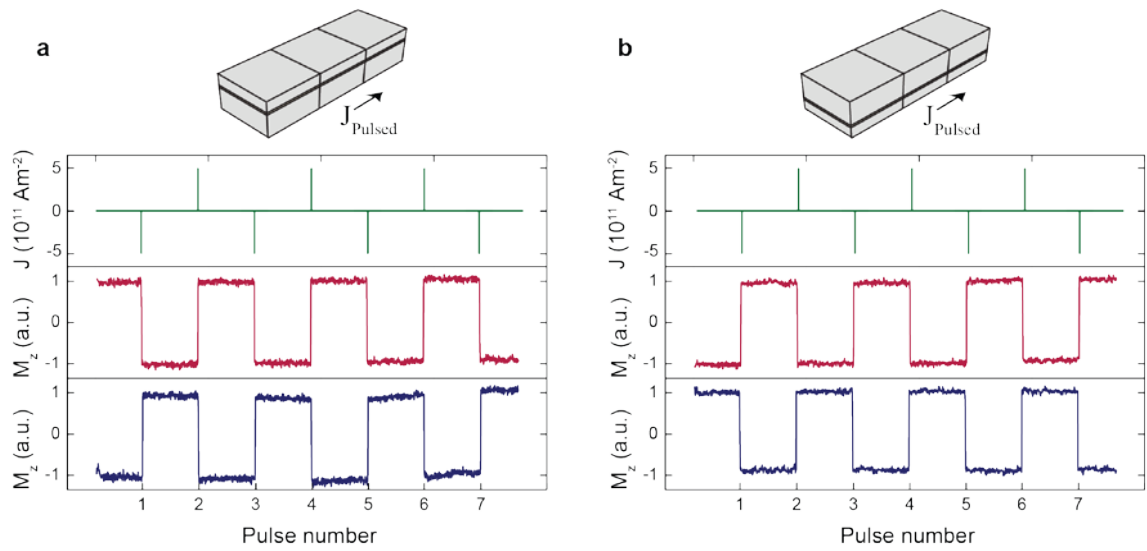

**Figure S4 | SHE-driven magnetization reversal**. **a** results for a Pt (4 nm) / Co (0.5 nm) / Pt (2 nm) and **b** results for a Pt (2 nm) / Co (0.5 nm) / Pt (4 nm) nanowire. The effect reverses, analogous to the reversal in sign of the domain wall depinning, as was shown in Fig. 3 of the main text. The red and blue lines denote the perpendicular component of the magnetization for $H_x = -20$ mT and $H_x = 20$ mT, respectively.